\documentclass[aps,twocolumn,groupedaddress,showpacs,floats,amssymb]{revtex4}
\usepackage[dvips]{graphicx}
\usepackage{bm}

\begin{document}

\title{On the Supersolid State of Matter}
\author{Nikolay Prokof'ev and Boris Svistunov}
\affiliation{Department of Physics, University of Massachusetts,
Amherst, MA 01003} \affiliation{Russian Research Center
``Kurchatov Institute'', 123182 Moscow }

\begin{abstract}
We prove that the necessary condition for a solid to be also a
superfluid is to have zero-point vacancies, or interstitial atoms,
or both, as an integral part of the ground state. As a
consequence, superfluidity is not possible in commensurate solids
which break continuous translation symmetry. We discuss recent
experiment by Kim and Chan [Nature, {\bf 427}, 225 (2004)] in the
context of this theorem, question its bulk supersolid
interpretation, and offer an alternative explanation in terms of
superfluid helium interfaces.
\end{abstract}


\maketitle Recent discovery by Kim and Chan \cite{chan,chan3} that
solid $^4$He samples have a non-classical moment of inertia (NCMI)
is a breakthrough result which has prompted renewed interest in
the supersolid (SFS) state of matter. Early theoretical work  by
Andreev and Lifshitz \cite{Andreev} and Chester \cite{Chester}
showed that solids may feature a Bose-Einstein condensate of
vacancies (or interstitial atoms) and thus possess superfluid (SF)
properties. One may consider their work as establishing  {\it
sufficient} conditions for SFS. It was natural then to interpret
mass decoupling in the torsion oscillator experiments as
originating from small (about $\sim 1\%$) concentration of
zero-point vacancies  \cite{chan}.

However, the overwhelming bulk of previous experimental work (for
review, see, e.g., \cite{meisel}) indicates that vacancies and
interstitials in $^4$He crystals are {\it activated} and their
concentration is negligible below $0.2~K$. The most recent study
\cite{Beamish} looked at the density variations of solid $^4$He
between two capacitor plates and did not reveal any presence of
vacancies.  To deal with these facts an idea was put forward that
strong exchange processes in quantum crystals may lead to
superfluidity even in the absence of zero-point defects
\cite{chan2,datta}. Mistakenly, this idea is attributed to
Leggett's paper \cite{leggett}. Leggett established a link between
the SF response and the {\it connectivity} of the groundstate
wavefunction and derived a rigorous formula for the upper bound on
the superfluid density, $\rho_s$.
Crystal defects and their relation to the connectivity was {\it
not} discussed in Ref.~\cite{leggett}. A separate issue is the
``extremely tentative" order-of-magnitude estimate $\rho_s \le
3\times 10^{-4}$ based on the exchange integral between $^3$He
atoms \cite{leggett}, which, apparently, caused the misleading
interpretation that interatomic exchange, on its own, may result
in SF in the absence of vacancies \cite{note2}.

The central point of the discussion to follow is to answer the
question whether superfluidity is possible in crystal structures
with the number of atoms being commensurate with the number of
lattice points and what are the necessary conditions for this to
happen.

Below we re-examine Leggett's work and show that it implies
vacancies and/or interstitial atoms as a {\it necessary} condition
for SFS in bosonic systems similar to $^4$He. Chester's ``final
speculation" that without them the solid state of matter is
insulating \cite{Chester} proves to be a theorem. We present an
alternative proof of the theorem using path-integral language in
which the presence of vacancies in the SF state is seen
explicitly. It also provides a simple picture showing why exchange
processes on their own do not lead to SF. Thirdly, we put forward
an argument based on the phase--particle-number uncertainty
relation \cite{anderson} which relates SF, compressibility of
pinned solids and vacancies (pinning is crucial to separate and
suppress the contribution to the compressibility coming from the
change of the lattice constant from that due to adding/removing
particles to/from the bulk \cite{note3}). The answer to the
central question is then that SFS groundstates in commensurate
solids have ``zero measure'' to be found in Nature, because they
require an exact symmetry between zero-point vacancies and
interstitials which is immediately broken by changing  system
parameters, e.g. pressure \cite{note4}. By excluding a bulk
supersolid interpretation of the Kim \& Chan results we are forced
to look for an alternative explanation of their data based on the
physics of disordered and frustrated $^4$He interfaces.

As shown by Feynman \cite{Feynman}, the groundstate of the bosonic
system has no zeros, $\Psi_G(x_1, x_2, \dots, x_N) \ne 0$.
Moreover, in superfluids $\Psi_G$ does not become macroscopically
small when one or several coordinates, say, $x_1, x_2, \dots x_m$,
are taken around the system while other coordinates are kept
fixed. This property (called connectivity by Leggett
\cite{leggett}) is key for superfluidity, and is just another way
of saying that topological off-diagonal long-range order (TODLRO)
is required for SF \cite{Kohn,leggett2,KT}. The requirement that
connected $\Psi_G$ be single-valued leads to the quantization of
circulation and thus stability of persistent currents in samples
with the cylindrical annulus geometry \cite{leggett}.

To illustrate the point, consider a one-dimensional system of two
identical bosons forming a bound (molecular) groundstate,
$\varphi_0(|r_1-r_2|)$, with localization length $l$. Naively, the
first rotating state of the molecule on a ring of large
circumference, $L \gg l$, is written as a product of the the plane
wave for the center of mass coordinate, $R=(r_1+r_2)/2$, times the
bound state: $\varphi_1=  e^{i 2 \pi R/L}\: \varphi_0(|r_1-r_2|)$.
This expression, however, is not single-valued, because if $r_1$
or $r_2$ is taken around the ring we get $\varphi_1 \to
-\varphi_1$. The correct solution is to replace $\varphi_0$ with
$\tilde{\varphi}_0$, which has a zero at $|r_1-r_2| \approx L/2$,
i.e. in the region where  $\varphi_0$  is exponentially small; at
distances $|r_1-r_2| \ll L$ the two functions are almost
identical, $|\tilde{\varphi}_0 | \approx \varphi_0$. Note, that
the energetic cost of creating a zero in this case is
exponentially small and vanishes in the limit of infinite system
size.


\begin{figure*}[tbp]
\begin{widetext}
\includegraphics[bb=20  0 560  765, angle=-90, width=0.68\columnwidth]{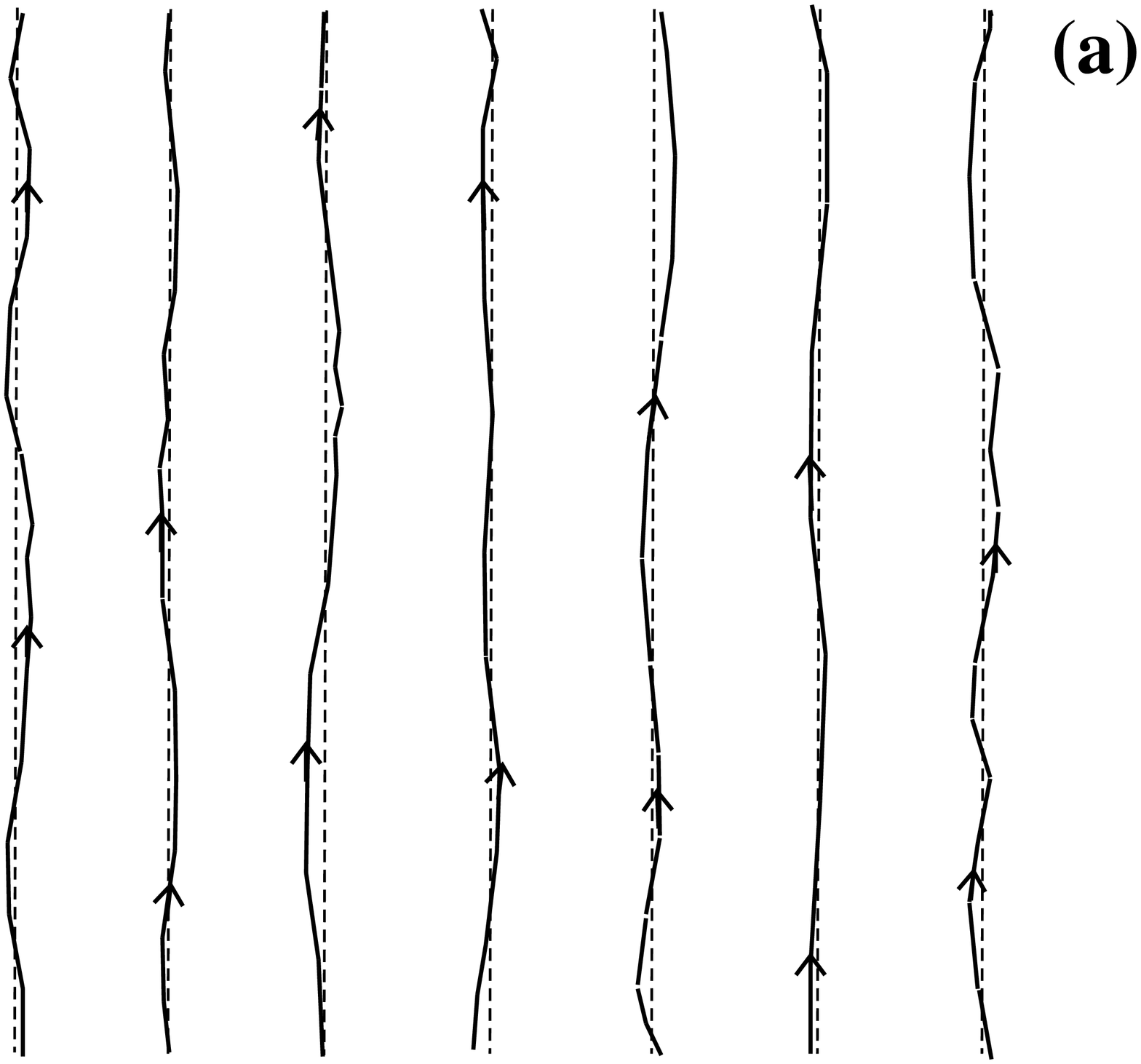}
\includegraphics[bb=20  25 560  750, angle=-90, width=0.68\columnwidth]{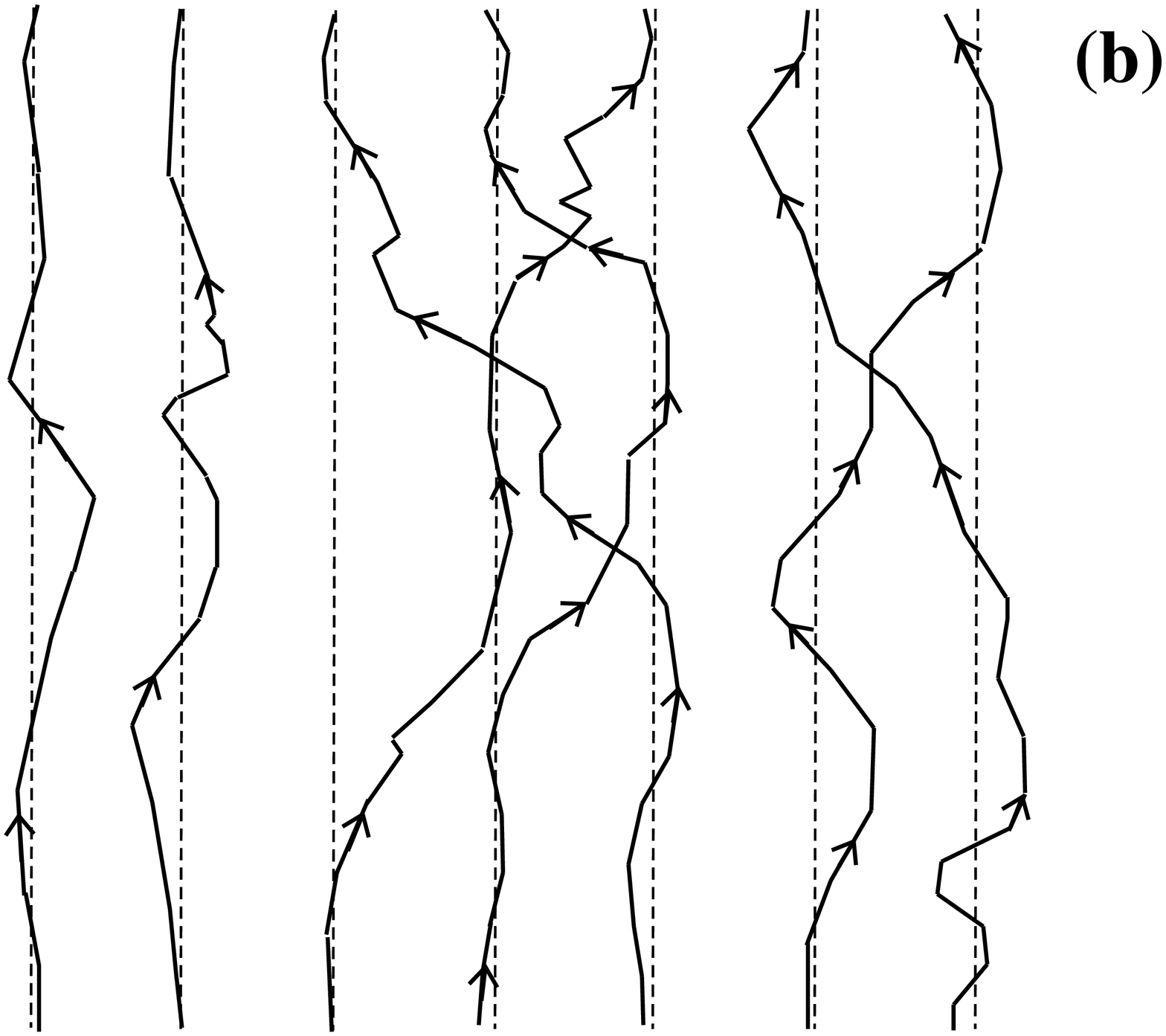}
\includegraphics[bb=20  25 560  750, angle=-90, width=0.68\columnwidth]{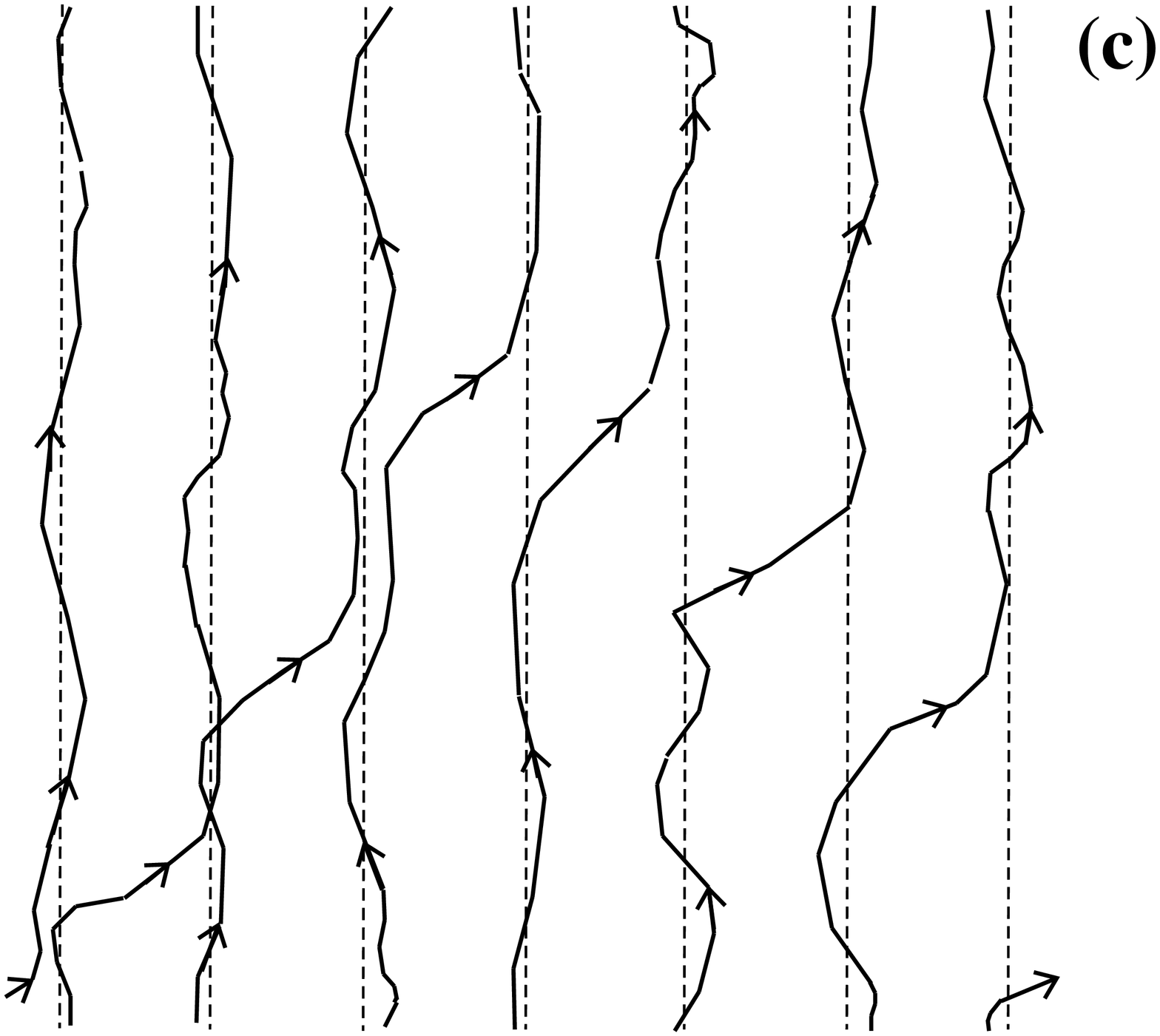}
\caption{ Particle worldlines in different crystals at low
temperature. The time axis is vertical. Dashed lines show the
equilibrium lattice points. (a) Nearly classical crystal at low
temperature; ${\bf M}=0$. (b) Insulating quantum crystal with
large zero-point fluctuations
    and frequent particle exchange processes; ${\bf M}=0$.
(c) Particle worldlines with non-zero winding number.
 }
\label{fig123}
\end{widetext}
\vspace*{-0.9cm}
\end{figure*}


The same considerations apply to the disconnected crystal state
consisting of $N$ bosons when $\Psi_G$ decays to the
macroscopically small value when any finite number of coordinates
are taken away from their typical positions in the crystal (other
coordinates are kept fixed) and moved around the system. The first
rotating state in the system with periodic boundary conditions can
be written as $\phi_1 =  e^{i 2 \pi R/L}\: \tilde{\varphi}_0 $
with $R=\sum_{i=1}^N r_i/N$ and $\tilde{\varphi}_0$ having zeros
in regions where the modulus of $\varphi_0$ is exponentially
suppressed and thus extra zeros do not cost finite (system size
independent) energy. Clearly, the phase gradient circulation of
$\phi_1$ is $\sim 1/N$, and this system will not show the NCMI
which is based on the impossibility of setting system in rotation
with a macroscopically small velocity or circulation. This should
be compared with the first rotating state of the single-atom
superfluid system, $\phi_1^{(SF)} =  e^{i 2 \pi \sum_{i=1}^{N} r_i
/L} \; \varphi_0 $ with $\varphi_0>0$ and phase gradient
circulation $2\pi$. Now, making zeros in connected $\varphi_0$ is
so costly energetically that the lowest energy state corresponds
to the relatively high kinetic energy of rotation. For
definiteness, we consider below only single-atomic superfluids,
but all considerations are readily generalized to the $m$-atomic
case. We have essentially reproduced above the Leggett's argument
that SF wavefunctions are necessarily connected.

By definition, $|\Psi_G(x_1, x_2, \dots , x_N) |^2$ is the
probability density to observe particles at the specified
positions. We fix all coordinates except one,  $x_1$, and observe
that in connected wavefunctions $|\Psi_G(x_1)|^2$ remains finite
when $x_1$ is taken arbitrary far from the initial position.
Formally, this property is identical to statistical properties of
atomic configurations in classical crystals at finite temperature
and was used by Chester to introduce vacancies in the ground
state.  This final correspondence was not elaborated in
Ref.~\cite{leggett}.

How do we ``visualize" vacancies/interstitials in the state of
identical particles with strong exchange, {\it especially when the
number of atoms coincides with the number of lattice points}? It
appears that the common perception is that such solids do not have
vacancies and interstitials, by definition, or else they are
indistinguishable from the standard zero-point fluctuations.
Imagine a solid sample pinned by an (arbitrarily weak) external
potential preventing it from moving as a whole. There is no
problem in identifying lattice points using the average particle
density profile $\rho (r)$ which is periodic in space. Consider
now some typical spatial configuration of particle positions along
with the lattice points and start the coarse-graining procedure of
``erasing'' the closest particle-lattice point pairs in the spirit
of the spatial renormalization group treatment. As we progress
towards mesoscopic length-scales all short-ranged zero-point
fluctuations of atoms away from lattice points will be erased from
the picture. The procedure continues until we have erased all
pairs with sizes much smaller than $L$ but much larger than all
microscopic scales. We say that the crystal state has no vacancies
and interstitials if the final coarse-grained configuration is
empty. On the other hand, if the coarse-grained configuration
still contains lattice points, or particles, or both, at arbitrary
large distances, we say that it has crystal defects in it. The
decimation procedure explains how vacancies are possible in
commensurate solids, and perfectly agrees with the conventional
view of classical crystals at finite temperature. For the
commensurate solid with connected $\Psi_G$ we may start with the
perfect-lattice configuration of particle coordinates and its
empty coarse-grained picture, and then move $x_1$ arbitrary
distance away to produce an image of the vacancy and interstitial.
This will not result in the exponential suppression of the
configuration probability (in fact, such configurations will
dominate in the normalization integral).

In the absence of exact interstitial/vacancy symmetry the
concentration of vacancies, $n_v$, and interstitials, $n_{\rm
int}$, in the supersolid will be different, e.g. $n_v>n_{\rm
int}$, since broken continuous symmetry allows production of
excess vacancies by making small changes in the lattice constant.
[This mechanism is is not available in discrete models, and then
$n_v=n_{\rm int}$ is possible.] In the translation invariant
system at T=0 one expects then $\rho_s =An_v$, just like in any
other interacting bosonic system.

Our second consideration is based on the path-integral formulation
of quantum statistics \cite{Feynman} in terms  of many-body
trajectories, $\{ x_i (\tau ) \}$, in imaginary time $\tau \in
[0,\beta ]$ with periodic boundary conditions $\{ x_i (\beta ) \}
= \{ x_i (0) \}$.
The most important superfluid characteristic of particle
trajectories, or world lines, is their winding numbers, $M^\alpha$,
$\alpha = 1,2,\dots , d$. We assume periodic boundary conditions
in space in all $d$-dimensions with linear sizes $L^\alpha =L$. To
determine $M^\alpha $ imagine a cross-section going through point
${\bf R}$ perpendicular to the direction $\alpha $ and count how
many times particles cross it from left to right, $k_-^\alpha $,
and from right to left, $k_+^\alpha $.
By definition, winding numbers
are $M^\alpha =   k_+^\alpha - k_-^\alpha$. They are independent
of the cross-section location ${\bf R}$ because trajectories are
continuous and periodic in imaginary time. The superfluid density
is then given by \cite{Ceperley}
\begin{equation}
\rho_s^{\alpha \gamma} =2mT L^{2-d} \: \langle \: M^\alpha
M^\gamma \rangle \;, \label{rho}
\end{equation}
where $m$ is the particle mass. In $d=3$, the superfluid density
is finite in the thermodynamic limit, $L \to \infty$, $T\to 0$,
and $T/L \to 0$, if the probability of having world lines with
non-zero winding numbers in the ground state is close to unity.

We now demonstrate that crystal states without zero-point
vacancies are described by world-line configurations with ${\bf
M}=0$, i.e. they are not superfluid. We start with the picture of
a perfect crystal with particles tightly localized around
equilibrium lattice points. In this state the trajectories are
nearly straight lines with ${\bf M}= 0$ as shown in Fig.~1(a).
When tunnelling exchange processes are added into the picture the
world lines are no longer in one-to-one correspondence with the
lattice points. The nature of the exchange process, however, is
such that when one particle leaves its equilibrium crystal point
$R_1$ and goes to point $R_2$, the other particle goes from  $R_2$
to $R_1$ (for pairwise exchange). Thus, the net current of
world lines through any cross-section remains strictly zero. The
same conclusion follows from the consideration of multiparticle
exchange events  \cite{ceperley2}, see Fig~1(b).

\begin{figure}[tbp]
\includegraphics[bb=0  100 600  700, angle=-90, width=0.6\columnwidth]{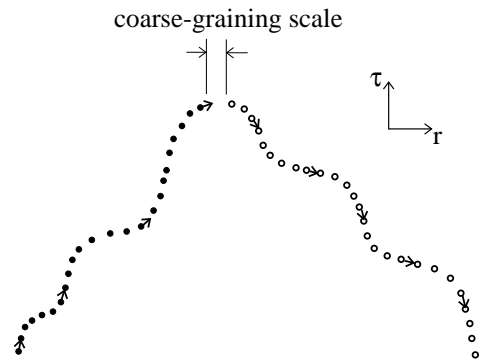}
\caption{ Evolution of the coarse-grained picture for the
trajectory with non-zero winding number similar to Fig.~1(c).
Filled and open  circles show particle and lattice site positions
correspondingly. Arrows indicate the direction of the particle
number current.
 }
\label{fig2}
\end{figure}

Consider now a world-line configuration with non-zero winding
number, Fig.~1(c). At any moment of imaginary time we consider the
spatial configuration of particle positions and lattice points and
apply the coarse-graining procedure discussed above. Again, all
short-lived and short-ranged exchange process and zero-point
fluctuations will be erased once we pass several atomic distances,
and  in the insulating state the ``movie'' of the coarse-grained
configuration evolution in time will show an empty ``screen'' for
Figs.~1(a,b). If ${\bf M} \ne 0$, as in Fig.~1(c), there is no way
for the renormalization procedure to erase all particles at all
moments in time since topologically winding numbers correspond to
particle trajectories moving continuously in the same spatial
direction and thus creating imbalance between particles and
lattice points at arbitrary large distances. After all other particles
are associated with lattice points and erased, the winding
trajectory describes an interstitial-vacancy pair which separates
over the distance of order $L$ and eventually makes a closed loop
around the system, see Fig.~\ref{fig2}. For the statistics of such
loops to give non-vanishing $\langle \: M^\alpha M^\gamma \rangle$
in the thermodynamic limit, they have to be typical in a given
lattice structure. In other words, zero-point vacancies and
interstitials are an integral part of the groundstate.

Our last consideration is based on the relation between the {\it
pinned} compressibility, $\kappa $, and zero-point vacancies.
Compressibility can be calculated through the energies of states
with one extra particle, $E_{N+1}$, and one extra vacancy,
$E_{N-1}$, as $\kappa = 1/V\Delta E$, where $V$ is the system
volume, and $\Delta E=E_{N+1}-E_{N-1}$. Incompressible states have
finite $\Delta E$. On another hand, in a macroscopic system
$\Delta E$ can be obtained by considering the energy increase by
creating an interstitial-vacancy pair with arbitrary large
separation between them (in pinned solids the notion of vacancy or
interstitial is rigorously defined within the coarse-graining
procedure as the presence or absence of perfect registry between
particles and unit cells). Crystals without zero-point defects are
gapped with respect to the interstitial-vacancy production
(otherwise these defects would be an essential part of the
groundstate) and thus are incompressible (if pinned) and {\it vice
versa}. One step further, this implies that superfluidity and
pinned compressibility come together and either one (including
long-wave acoustic properties with additional sound mode) can be
used for the detection of the SFS state experimentally. This
conclusion is in line with the famous uncertainty relation
\cite{anderson} between the phase of the superfluid order
parameter, $\phi$, and particle number, $\Delta \phi \Delta N \ge
1/2$. Because of this relation, one may not introduce a well
defined phase field for the incompressible state of matter which
tends to {\it completely} suppress particle number fluctuations.

We have little doubt that large activation energies for vacancies
and interstitials in $^4$He measured down to $\sim1~K$
temperatures \cite{meisel} will not radically change to near zero
at lower temperatures, and that helium is a commensurate solid at
$T=0$. Since it has no symmetry between the vacancies and
interstitials, we conclude that there are no zero-point vacancies
in bulk solid $^4$He. By excluding superflow through the crystal
bulk we are forced to look more closely at the superfluid
properties of disordered helium-substrate layers and frustrated
interfaces between helium micro-crystallites.

There are indications of strong disorder in experimental system of
Ref.~\cite{chan}. The dependence of the superfluid density on
reduced temperature parameter $t=(T_c-T)/T_c$ has little to do
with the expected bulk superfluidity $t^{0.671}$ dependence.
Instead, $\rho_s$ appears to vanish at $T_c$ with zero derivative.
Such a behavior can be modelled by a broad distribution of
transition temperatures in the heterogeneous sample. This
observation correlates with the gradual decrease of the decoupled
mass with the increase of the torsion oscillator amplitude by
orders of magnitude. Let us assume that a sample consists of
micro-crystallites of linear size $D$ with superfluid interfaces
of typical thickness $d$ between them. The superlfuid fraction may
be estimated then as $\rho_s/\rho \sim d/D$. To have $\sim 1\%$ of
the superfluid mass coming from interfaces with $d\sim 10~\AA$ one
will need crystallite sizes about a fraction of a micron. The
variety of interfaces with different crystallographic indices
provides a broad distribution of transition temperatures.

One of the experimental mysteries is extreme sensitivity to the
addition of $^3$He impurities at the level of $n_{\rm im} \sim 100
ppm$. To minimize kinetic energy, light $^3$He atoms are likely to
end up at frustrated interfaces, and then at the edges where
different interfaces meet. This may increase $^3$He edge {\it vs }
bulk concentration by a factor as large as $(D/d)^2$ and produce
$n_{\rm im}^{({\rm edge})} \sim 1$. This will have a profound
effect on the edge-connected interface superfluidity (edges then
act as a disordered two-dimensional network of Josephson junctions).

We are not aware of any systematic study what are the properties
of interfaces between the $^4$He micro-crystals and helium $^4$He
crystals on disordered substrates at elevated pressures. Model
simulations of domain wall boundaries in the checkerboard solid
(obtained for interacting lattice bosons at half-integer filling
factor) show that they remain superfluid deep into the bulk solid
phase \cite{note6}. In the outlined picture three predictions are
certain: (i) the superfluid fraction must strongly depend on the
crystal growth process, (ii) the amount of $^3$He
sufficient to suppress superfluidity scales as $n_{\rm im} \propto
\rho_s^2$, and (iii) transition temperatures on interfaces do 
not depend on $D$ and $\rho_s (T=0)$.

We are grateful to M. Chan, R. Guyer, R. Hallock, Yu. Kagan, A.
Leggett, J. Machta, and W. Mullin for stimulating discussions. The
research was supported by NSF under grant No. PHY-0426881 and by
NASA under grant NAG32870.


\begin{thebibliography}{99}
\bibitem{note2} Even if vacancies
make up $\sim 1\%$ of lattice cites the upper bound is increased
by orders of magnitude due to much larger particle-vacancy
exchange integral.
\bibitem{note3} Qualitatively, pinned crystals are equivalent to lattice
bosonic systems.
\bibitem{note4} ``Zero-measure" argument excludes systems with
explicitly broken translation symmetry, e.g. by external periodic
potential, where commensurability may be enforced by hand and the
lattice constant may not be changed continuously.
\end{thebibliography}
\end{document}